\documentclass[pr,twocolumn,superscriptaddress]{revtex4-1}
\usepackage[english]{babel}
\usepackage{amssymb,amsmath}
\usepackage{dsfont}
\usepackage{gensymb}
\usepackage{color}
\usepackage{epsfig}
\usepackage{graphicx}
\usepackage[caption=false]{subfig}

\begin{document}

\title{Enhanced thermoelectric figure of merit in polycrystalline carbon nanostructures}

\author{Thomas~Lehmann}
\affiliation{Institute for Materials Science and Max Bergmann Center of Biomaterials, TU Dresden, 01062 Dresden, Germany}
\affiliation{Dresden Center for Computational Materials Science (DCMS), TU Dresden, 01062 Dresden, Germany}
\author{Dmitry\,A.~Ryndyk}
\affiliation{Institute for Materials Science and Max Bergmann Center of Biomaterials, TU Dresden, 01062 Dresden, Germany}
\affiliation{Dresden Center for Computational Materials Science (DCMS), TU Dresden, 01062 Dresden, Germany}
\affiliation{Center for Advancing Electronics Dresden, TU Dresden, 01062 Dresden, Germany}
\author{Gianaurelio~Cuniberti}
\affiliation{Institute for Materials Science and Max Bergmann Center of Biomaterials, TU Dresden, 01062 Dresden, Germany}
\affiliation{Dresden Center for Computational Materials Science (DCMS), TU Dresden, 01062 Dresden, Germany}
\affiliation{Center for Advancing Electronics Dresden, TU Dresden, 01062 Dresden, Germany}
\date{\today}

\begin{abstract}

Grain boundaries are commonly observed in carbon nanostructures, but their influence on thermal and electric properties are still not completely understood. Using a combined approach of density functional tight-binding theory and non-equilibrium Green functions we investigate electron and phonon transport in carbon based systems. In this work, quantum transport and thermoelectric properties are summarized for graphene sheets, graphene nanoribbons and carbon nanotubes with a variety of grain boundary types in a wide temperature range. Motivated by previous findings that disorder scatters phonons more effectively than electrons, a significant improvement in the thermoelectric performance for polycrystalline systems is expected. As the effect is marginally sensitive to the grain boundary type, we demonstrate that grain boundaries are a viable tool to greatly enhance the figure of merit, paving the way for the design of new thermoelectric materials.

\end{abstract}

\maketitle

\section{Introduction}
Graphene, the building block for various carbon nanostructures of all other dimensionalities, is the subject of exceptional scientific interest in the recent years \cite{Novoselov:2004,Geim:2007}. Today, graphene sheets, graphene nanoribbons and carbon nanotubes are considered as promising candidates in the vast field of electronic, thermal and thermoelectric applications.
However, macroscopic samples of those structures will most certainly appear in polycrystalline form, an issue which is neglected in most studies. Polycrystallinity is an inherent structural impurity and very challenging to avoid. It is caused by growth kinetics or substrate imperfections, resulting in different graphene domains with a variety of crystallographic orientations \cite{Loginova:2009}. The common domain interface, the grain boundary (GB), can be regarded as an one-dimensional array of dislocations. In an infinite two-dimensional graphene sheet, this dislocation array can be approximated as a linear periodic array \cite{Shockley:1950}, whereas its periodicity or Burgers vector depends on the lattice mismatch. Besides unintentional polycrystallinity of the system as an inherent property, structuring of well-defined line defects has been demonstrated in graphene and opens new possibilities \cite{Lahiri:2010aa}.
The resulting structures have been addressed both from experimental and theory groups and can be classified by either their relative orientation angle or lattice mismatch \cite{Lahiri:2010aa,Loginova:2009,Yazyev:2010,Yazyev:2010a, Pop:2012}. Despite a large number of experimental observations, the implications on transport properties are not yet completely understood. 
It has been shown that heat flow in polycrystalline graphene depends on the specific GB structure \cite{Serov:2013aa} by effectively scattering phonons at the interface. These findings are encouraging to investigate the issue of thermoelectrics, which complements the recent studies focussing on either electronic or thermal transport \cite{Sevincli:2013,Lehmann:2013}. Additionally, we widen the scope of interest beyond two-dimensional graphene by including quasi-1D carbon allotropes, like graphene nanoribbons (GNR) and carbon nanotubes (CNT) in our studies. In fact, we show that GBs can not only tune a transport gap controlling charge currents \cite{Wang:2012,Koepke:2013}, but also significantly increase the thermoelectric figure of merit. Combined with the exceptional charge transport in carbon nanostructures, GBs seem to evolve as promising candidates in scattering phonons to suppress the high thermal conductivity, which was found to reach up to 5000~Wm$^{-1}$K$^{-1}$ for free-standing single-layer graphene \cite{Balandin:2008}.

This work presents a summary of electronic and thermal transport properties in one-dimensional and two-dimensional carbon nanostructures with GBs in order to study the use of polycrystalline structures in thermoelectric materials. The paper is structured as follows. After a short introduction to motivate the topic in Sec. I, we will give an overview of the various structures and the theoretical framework for charge and thermal transport calculation in the next section of this paper, Sec. II. Results will be discussed in Sec. III for each of the systems separately, i.e., graphene, GNRs and CNTs, and we will summarize and conclude subsequently.

\section{Model and Theory}
\label{sec:model}
The polycrystalline nanostructures have been constructed by joining two subsystems of different chiralities or crystal orientations. Based on those initial geometries, the minimum energy configurations of the structures were then determined using molecular dynamics simulation followed by density functional geometry optimization until the force convergence criterium of 0.02~eV\AA$^{-1}$ was met. Hydrogen saturation of eventually di-coordinated carbons was neglected. In our analysis, the variety of GBs in graphene is limited by the need for a reasonable periodicity along the interface, otherwise the system size gets too large for \textit{ab initio} calculations. This limitation is not apparent in the quasi-1D systems of GNRs and CNTs, but the choice of diverse chiralities is restricted by roughly matching the CNT diameters.

For the analysis of quantum transport characteristics, the Green function formalism has been applied in combination with the efficient density functional tight-binding approach (DFTB), as implemented in DFTB+ \cite{Aradi:2007,Pecchia:2008}. This methodology allows the calculation of electron and phonon transport properties in the ballistic transport regime based on the same theoretical footing \cite{Mingo:2003,Mingo:2007}. Transmission spectras are obtained in the first place, which enable the calculation of electron conductance, current-voltage characteristics, thermopower, thermal conductance and thermoelectric figure of merit.
The here applied ballistic transport model neglects phonon-phonon and electron-phonon interactions, but the high intrinsic mean free paths for electrons and phonons at room temperature in carbon systems \cite{Pop:2012} validate this approximation. Furthermore, the model is superior to classical molecular dynamics approaches for thermal transport, as they lack quantum features like Bose-Einstein statistics. The high Debye temperature of about 2100~K in graphene systems \cite{Pop:2012} necessitates quantum calculations. 
The electron transmission function $T_{\mathrm{el}}(E)$ was obtained by the standard single-particle Green function formalism \cite{Datta95book}:
\begin{eqnarray}
  \label{eq:Te}
  T_{\mathrm{el}}(E) &= \textrm{Tr}\left[\hat{\Gamma}^{\mathrm{el}}_{\mathrm{L}} \hat{G} \hat{\Gamma}^{\mathrm{el}}_{\mathrm{R}} \hat{G^\dag}\right],\\
  \hat{G}(E) &= \left[ E\hat{S}-\hat{H}\right]^{-1}.
\end{eqnarray}
The Hamiltonian $\hat{H}$ of the scattering region has been calculated using DFTB and semi-infinite unperturbed leads are assumed and incorporated by self-energy terms using the decimation technique by L\`{o}pez Sancho \textit{et al.} \cite{Sancho:1985}.
Phonon transmission spectra $T_{\mathrm{ph}}(\omega)$ were calculated analogously based on the atomistic Green function method \cite{Mingo:2007},
\begin{eqnarray}
  \label{eq:Tp}
  T_{\mathrm{ph}}(\omega) &= \textrm{Tr}\left[\hat{\Gamma}^{\mathrm{ph}}_{\mathrm{L}} \hat{\mathcal{G}} \hat{\Gamma}^{\mathrm{ph}}_{\mathrm{R}} \hat{\mathcal{G}^\dag}\right], \\
  \hat{\mathcal{G}}(\omega) &= \left[ \omega^2\hat{\mathds{1}}-\hat{D}\right]^{-1},
  \label{eq:8}
\end{eqnarray}
using the dynamical matrix $\hat{D}$ obtained from the mass-weighted force constant matrix, which can be evaluated from the second derivatives of the lattice potential energy with respect to spatial displacements $\hat{K}=\{k_{ij}\}=\partial^2 U/\partial x_i \partial x_j$. This Hessian matrix has also been obtained by DFTB calculations and, in analogy of the electronic counter part, the leads account for self-energy terms calculated with the decimation technique. For further details on the Green function formalism for electrons and phonons we refer to the corresponding literature or previous publications \cite{Sevincli:2013,Lehmann:2013}.

For a shorthand notation for several properties, we introduce the Onsager coefficients 
\begin{equation}
  \label{eq:1}
  L_n(T)=\int{\left(E-E_{\mathrm{F}}\right)^n \left(-\frac{\textrm{d}f_{\mathrm{F}}(E,T)}{\textrm{d}E}\right)T_{\mathrm{el}}(E)\textrm{d}E},
\end{equation}
with $n \in \mathds{N}_0$, Fermi energy $E_{\mathrm{F}}$ and Fermi-Dirac distribution $f_{\mathrm{F}}(E,T)$.
Based on the electron transmission function $T_{\mathrm{el}}$ we calculate the electric conductance according to the Landauer formula $\sigma = \frac{2e^2}{h}T$, the temperature dependent Seebeck coefficient or thermopower
\begin{equation}
  \label{eq:2}
  S(T) = -\frac{1}{eT}\frac{L_1}{L_0},
\end{equation}
and the current-voltage characteristics for a source-drain voltage $V$ applied between the two contacts
\begin{equation}
  \label{eq:3}
  I(V) = \frac{2e}{h} \int{ T_{\mathrm{el}}(E) \left[f_{\mathrm{F}}(E,T)-f_{\mathrm{F}}(E+eV,T)\right]\textrm{d}E}.
\end{equation}
The thermal conductance $\kappa$, consisting of an electronic contribution $\kappa_{\mathrm{el}}$ and the thermal conductance of the lattice $\kappa_{\mathrm{ph}}$, can then obtained by the phonon transmission function $T_{\mathrm{ph}}$
\begin{equation}
  \label{eq:4}
\kappa=\underbrace{ \frac{2}{hT} \left(L_2-L_1^2/L_0\right)}_{\kappa_{\mathrm{el}}}+\underbrace{\frac{\hbar}{2\pi}\int{\omega T_{\mathrm{ph}}(\omega) \frac{\textrm{d}f_{\mathrm{B}}(\omega,T)}{\textrm{d}T} \textrm{d} \omega}}_{\kappa_{\mathrm{ph}}},
\end{equation}
where $f_{\mathrm{B}}(\omega,T)$ is the Bose-Einstein distribution function. In general, heat flow is generated by lattice vibrations and free conduction electrons. However, in carbon nanostructures the electronic thermal transport is very limited \cite{Jiang:2011} and accounts only for a few per mille of the total conductance. 

Finally, we compute the thermoelectric figure of merit
\begin{equation}
  \label{eq:5}
  ZT = \frac{\sigma S^2 T}{\kappa} = \frac{1}{(L_0L_2/L_1^2)-1}\frac{\kappa_{\mathrm{el}}}{\kappa},
\end{equation}
which comprises of all previous quantities and characterizes the efficiency of the thermoelectric effect in the system.

\section{Discussion of Results}
\label{sec:results}

\subsection{Graphene}
\label{sec:graphene}
\begin{figure}

  \subfloat[Examples of graphene GBs. The symmetric structure $(2,1)|(2,1)$ corresponds to a class-I GB, whereas the asymmetric GB $(5,3)|(7,0)$ is class II. The corresponding matching vectors $(n_{\mathrm{L}},m_{\mathrm{L}})$, $(n_{\mathrm{R}},m_{\mathrm{R}})$ are shown in black.]{
     \includegraphics{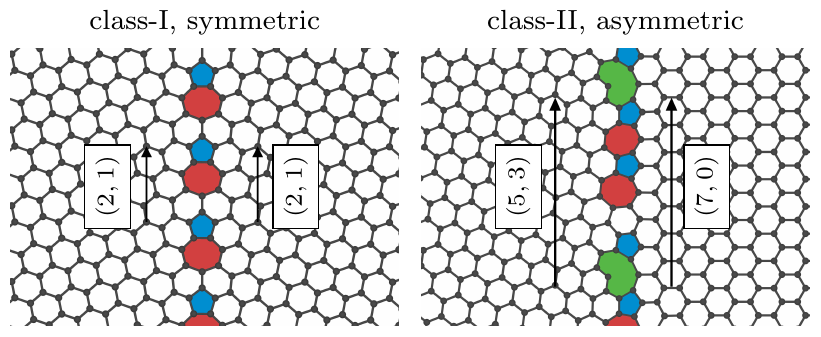}
    \label{fig:graphene}}
  
  \subfloat[Examples of GBs in GNR structures, obtained by joining two $\approx 1.7~nm$ wide GNRs with different inclination angles $\Theta$.]{
    \includegraphics{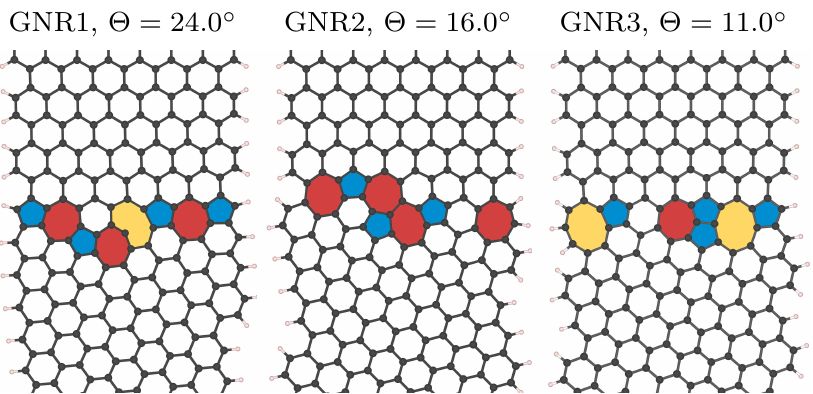}
    \label{fig:GNR}}

  \subfloat[Examples of GBs in CNT structures, obtained by joining two nanotubes with a similar diameter of about 0.55~nm. The chirality indices are given for each pair of CNTs.]{
    \includegraphics{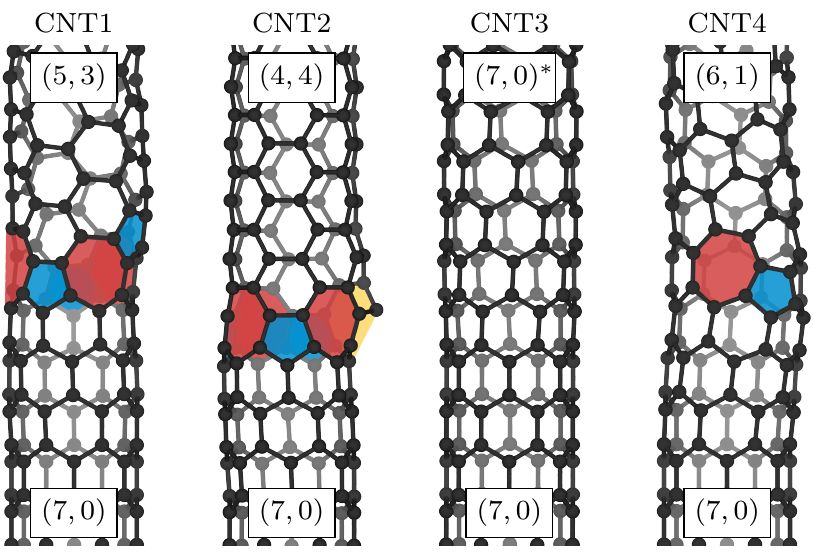}
    \label{fig:CNT}}
  \caption{(Color online) Examples of polycrystalline carbon nanostructures used for transport calculation. The GBs, obtained by DFT geometry optimization, exhibit pentagon (blue), heptagon (red), octagon (yellow) and nonagon defects (green).}
  \label{fig:geom}
\end{figure}

\begin{figure*}
    \includegraphics{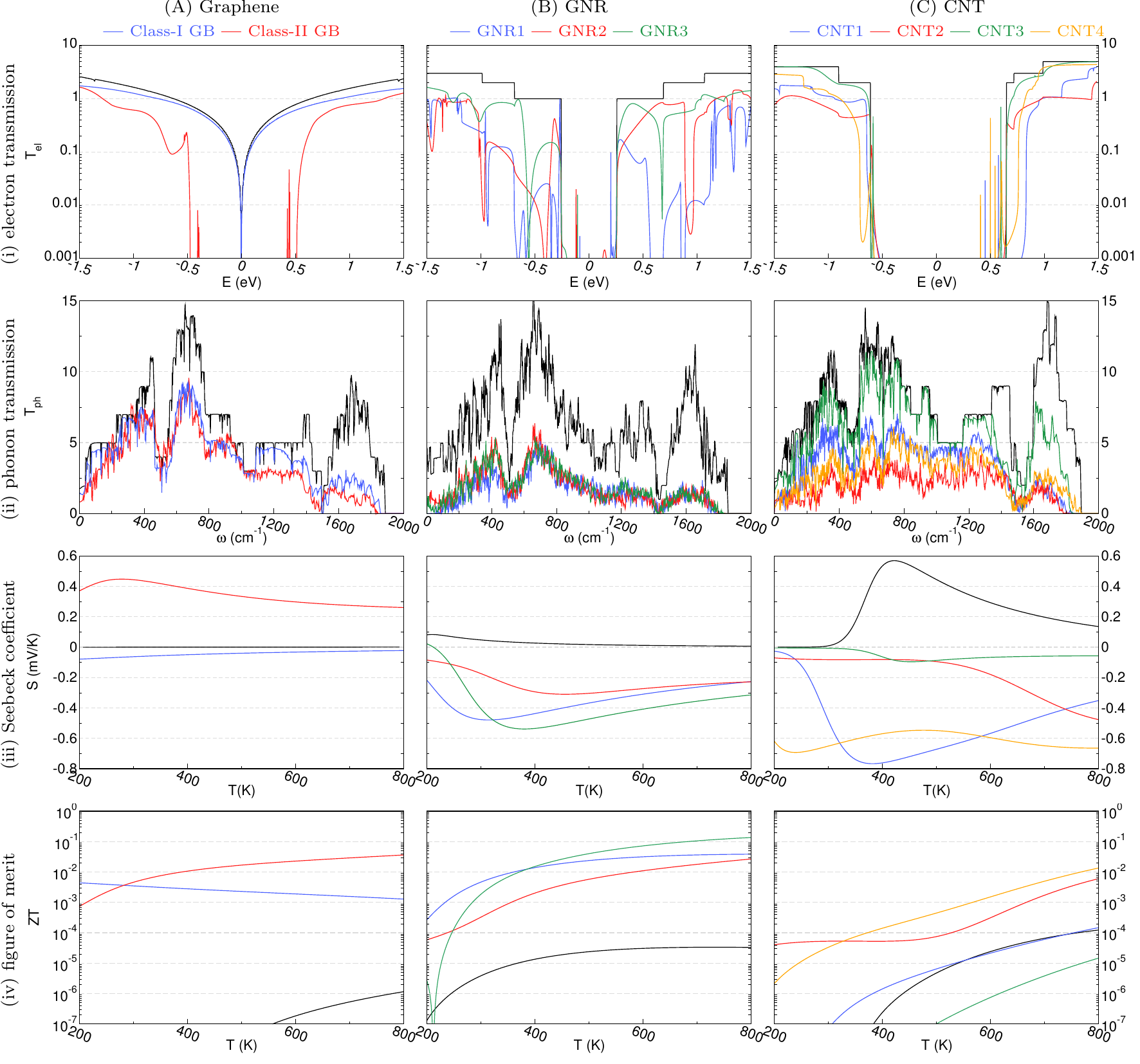}

  \caption{(Color online) Calculated electron transmission (i) and phonon transmission (ii) at room temperature, as well as the Seebeck coefficient (iii) and thermoelectric figure of merit (iv) as a function of temperature for graphene (A), graphene nanoribbons (B) and carbon nanotubes (C) with different GB types, see Fig. \ref{fig:geom}. Black curves represent the monocrystalline system for reference.}
  \label{fig:thermal}
\end{figure*}

The GBs in graphene sheets can be approximated as linear periodic arrays of dislocations \cite{Shockley:1950}.
We are using a GB classification in graphene proposed by Yazyev \textit{et al.} \cite{Yazyev:2010}, separating into two classes corresponding to their matching vectors $(n_{\mathrm{L}},m_{\mathrm{L}})$ and $(n_{\mathrm{R}},m_{\mathrm{R}})$. If exactly one matching vector fulfills the criterion $(n-m)=3q~(q\in\mathbb{Z})$, the GB is of class-II type. Otherwise it belongs to class-I. Due to a misalignment of allowed momentum-energy manifolds, class-II type boundaries introduce a transport gap, which can be approximated by $E_{\mathrm{G}}=hv_{\mathrm{F}}/3d = \frac{1.38~ \textrm{eV}}{d~\textrm{(nm)}}$, solely depending on the periodicity $d$ \cite{Yazyev:2010}. The distinct behavior of both classes can be explained by transverse momentum conservation at the interface and an effective rotation of the Brillouin zone for the charge carriers passing the interface.
We concentrate on two examples, one symmetric (class-I) and one asymmetric (class-II) GB, see Fig. \ref{fig:graphene}. For reference, the properties of an unperturbed graphene sheet are calculated. As expected, the asymmetric class-II GB exhibits an energy gap of about 1~eV, whereas the transmission spectrum of the symmetric class-I GB in Fig. \ref{fig:thermal}~A(i) is very similar to the pristine sheet. Phonon transmission shows a weak dependence on the GB type, but a slightly stronger phonon scattering can be identified for class-II GB with differences most pronounced at very low ($<200~$cm$^{-1}$) and high phonon energies ($>1000~$cm$^{-1}$), see Fig. \ref{fig:thermal}~A(ii). This can be attributed to a stronger lattice deformation and buckling along the interface. The introduced asymmetry of electron and hole transmission leads to a separation of charge carriers, improving the thermopower $S$ in those systems over ideal graphene. Electron holes account for the Seebeck effect in class-II GBs, and therefore $S>0$, and give rise to an improved thermopower compared to the electron-dominated symmetric GBs. For the thermoelectric figure of merit $ZT$ an enhancement by over three orders of magnitude in Fig. \ref{fig:thermal}~A(iv) is expected above room temperature for both types. Interestingly, both GB types perform equally well, as the lower Seebeck coefficient $S\approx 0.05$~mVK$^{-1}$ of the class-I GB is one order of magnitude smaller than for class-II type interfaces, $S\approx 0.4$~mVK$^{-1}$ in Fig. \ref{fig:thermal}~A(iii), but gets compensated by an improved electron conductance. 

\subsection{Graphene nanoribbons}
\label{sec:gnr}
As the gapless nature of graphene has negative effects on its thermoelectric properties, spatial confinement leads to an intrinsic band gap which depends on size and edge structure. By parallel cutting polycrystalline graphene sheets into ribbons, one expects structures similar to those shown in Fig. \ref{fig:GNR}, i.e., at least two graphene nanoribbons of different crystallographic orientation with a GB at the interface. Compared to graphene sheets, those structures omit periodicity parallel to the interface and reveal hydrogen termination at the edges to saturate dangling bonds and to remove particular edge state effects. The interface in between constitutes of an array of dislocations, either pentagons, heptagons, or octagons. For one side of the system, an about 1.7~nm wide armchair GNR with 15 dimer lines in width was chosen. The ribbon across is rotated respectively by an angle $\Theta$, ranging from 11\degree to 24\degree. The angles are chosen in a way that the periodic supercell of the lead does not get unnecessarily large. By cutting the ribbon to the same width as the right part one produces irregular but periodic edge geometries. As a reference, the results for the unperturbed armchair ribbon are shown and, like all armchair terminated ribbons, it is semiconducting with a band gap of about 0.5~eV. As shown in Fig. \ref{fig:thermal}~B(i), the suppression of electron transmission is strongest in GNR1, which may be explained by the non-interrupted dislocation array, compared to the other configurations GNR2 and GNR3, where one hexagon in the interface is preserved. One also notices the electron-hole symmetry breaking, with an improved transmission for low energy electrons compared to respective holes. Interestingly, the effect of different GBs are not apparent in the thermal conductance. As expected, phonons get scattered at the interface, effectively lowering the transmission by roughly 50\%, but the spectrum is mainly independent on the exact geometry, see Fig. \ref{fig:thermal}~B(ii). This results in about one tenth of the thermal conductance of an ideal ribbon. Effects on the electronic and thermal properties combined give rise to a significantly improved thermoelectric figure of merit, shown in Fig. \ref{fig:thermal}~B(iv). At room temperature, we calculated an increase by three to four orders of magnitude, topping at around $ZT\approx 0.1$ for GNR3 at 700~K. The zero in the $ZT$ graph of GNR3 can be associated to a change of the majority charge carrier type from holes to electrons with increasing temperature, implying a sign change of the Seebeck coefficient $S$ at 210~K. In fact, $S$ is negative with values around $S\approx -0.4$~mVK$^{-1}$ for all configurations, in contrast to $S>0$ in the ideal GNR. Those findings are particularly promising as all three samples show the same qualitative performance, suggesting a robust and efficient method for enhancing the thermoelectric effect.

\subsection{Carbon nanotubes}
\label{sec:cnt}

Next to the planar carbon structures, their rolled-up form is equally promising. Here we will shed light on the electric and thermal properties of carbon nanotube heterojunctions. Carbon nanotubes of different chirality but similar diameter can form a junction by exhibiting a couple of dislocation defects, see Fig. \ref{fig:CNT}. Such structures can evolve if two individual tubes eventually grow together, or by a change in the growth parameters \cite{Yao:1999,Yao:2007,Bandow:1998,Doorn:2005}. First interest in application of intra molecular nanotube junctions is the use as heterojunction diodes \cite{Treboux:1999}.
We studied four different CNT heterojunctions, each of them with a diameter of about 0.55~nm. In all samples one part is made of a (7,0)-CNT, which is complemented by a second CNT with aligned tube axis, see Fig. \ref{fig:CNT}. Kinked carbon nanotube junctions are not considered in this work. Notice, that the structure CNT3 in Fig. \ref{fig:CNT} shows no chirality change, but a twist around the tube axis. The twisting angle is given by (or multiples of) $\phi = \frac{90 \degree}{7} \approx 12.8\degree$ due to the seven-fold symmetry. This twist deforms the lattice and increases stress at the interface, but does not lead to defect formation or reconstruction. In the remaining three structures CNT1, CNT2 and CNT4, the chirality indices (5,3), (4,4) and (6,1) have been chosen, all yielding a diameter close to its counterpart of 0.55~nm.
Primarily one observes pentagon-heptagon pairs along the circumference, and one octagon in CNT2. The properties of the ideal and semiconducting (7,0)-CNT are calculated as a reference. Fig. \ref{fig:thermal}~C(i) shows the energy gap in the electron transmission appearing in all of the different heterojunctions. Most notably, the twisted carbon nanotube, CNT3, shows only little change in the electron transmission where the slight lattice perturbations smear out the perfect transmission steps of an ideal nanotube. Electrons in CNT1 and CNT4 are suppressed equally and transmission is roughly cut in half. The heterojunction in CNT2 shows the strongest scattering of electrons, effectively lowering electron conductance. We attribute this to the strong deformation caused by the octagon defect. A similar picture can be drawn for phonon transport, see Fig. \ref{fig:thermal}~C(ii). As no defects are formed in CNT3, lattice vibrations are weakly affected and scattering happens only for high frequency phonons. But for heterojunctions of different chiralities, phonon transmission gets significantly suppressed at all frequencies. 
Interestingly, CNT4 with only one defect pair is superior to structure CNT1 in blocking phonons and the defect configuration in CNT2 scatters strongest. Consequently, thermal conductance is decreased up to a factor of five for the CNT2 structures compared to the ideal nanotube. Quantitatively, this is very similar to the case of GNRs discussed above. In terms of figure of merit, the thermoelectric properties can be significantly improved but strongly depend on the exact structure, see Fig. \ref{fig:thermal}~C(iv). At room temperature, a gain in calculated $ZT$ as high as two orders of magnitude for the structures CNT1 and CNT2 can be reported. Note that the twisted structure without chirality change, CNT3, shows a decreased $ZT$ by a factor of ten. This is due to the loss of electron-hole asymmetry as the slight distortions smear out the sharp band edges. Thereby the Seebeck effect is strongly reduced. With the right choice of CNT chiralities, the Seebeck effect can be increased to values around $S\approx -0.6$~mVK$^{-1}$ for a broad temperature range and a substantial enhancement of $ZT$ is expected.

\section{Conclusions}
\label{sec:conclusions}

A material with high thermoelectric efficiency has to show properties of an electron crystal and phonon glass at the same time, i.e., blocking thermal transport while maintaining electric current. Our results show that lattice imperfections forming at the interface of two carbon systems with different crystallographic orientations scatter heat stronger than charge carriers. In studying the effect of various GB types on the electron and phonon transport properties of polycrystalline carbon nanostructures, we predict improved thermoelectric properties. The odd-membered rings at the boundary break the bipartite symmetry of the lattice resulting in an electron-hole asymmetry in the electron transmission spectrum. This charge carrier separation is advantageous for the thermoelectric effect, increasing the thermopower. This, combined with suppressed thermal transport due to phonon scattering at atomic dislocations, in particular for mid to high energy phonons, generates a substantial improvement in the figure of merit. Our calculations show for nearly all tested configurations an enhancement of several orders of magnitude in $ZT$ at room temperature. The low sensitivity of this effect on the GB type heavily facilitates any experimental realization. We also note that there is even more room for improvement of $ZT$ by shifting the chemical potential. This opens up promising prospects for the use of polycrystalline carbon nanostructures in thermoelectric applications.

\section{Acknowledgments}
We thank Dr. Arezoo Dianat and Dr. Frank Ortmann for fruitful discussions. This work is funded by the EU FP7 projects NanoCaTe and PAMS. It is partly supported by the German Research Foundation (DFG) within the priority program Nanostructured Thermoelectrics (SPP 1386) and the Cluster of Excellence "Center for Advancing Electronics Dresden". We acknowledge the cooperation with the National Science Center of Poland in the research program Harmonia (DEC-2013/10/M/ST3/00488) and the Center for Information Services and High Performance Computing (ZIH) at TU Dresden for computational resources.

\bibliography{bibliography.bib}
\end{document}